\documentclass[12pt]{article}

\usepackage{eniac2023}
\usepackage{graphicx,url}
\usepackage[utf8]{inputenc}
\usepackage[brazil]{babel}

\title{A Deep Learning Application for Psoriasis Detection}

\author{Anna Milani\inst{1}, Fábio S. da Silva\inst{2}, Elloá B. Guedes\inst{2}, Ricardo Rios\inst{1}}

\address{Escola Superior de Tecnologia \textemdash~ Universidade do Estado do Amazonas
\\ Manaus, AM, Brasil
\nextinstitute
Laboratório de Sistemas Inteligentes \textemdash~ Universidade do Estado do Amazonas
\\ Manaus, AM, Brasil
  \email{\{aavm.snf19, fssilva, ebgcosta, rrios\}}@uea.edu.br
}

\begin{document}

\maketitle

\begin{abstract}
  In this paper a comparative study of the performance of three Convolutional Neural Network models, ResNet50, \textit{Inception v3} and VGG19 for classification of skin images with lesions affected by psoriasis is presented. The images used for training and validation of the models were obtained from specialized platforms. Some techniques were used to adjust the evaluation metrics of the neural networks. The results found suggest the model \textit{Inception v3} as a valuable tool for supporting the diagnosis of psoriasis. This is due to its satisfactory performance with respect to accuracy and F1-Score (97.5\% ${\pm}$ 0.2).
\end{abstract}

\begin{resumo}
  Nesse artigo é apresentado um estudo comparativo do desempenho de três modelos de Redes Neurais Convolucionais, ResNet50, \textit{Inception v3} e VGG19 para classificação de imagens de pele com lesões afetadas por psoríase. As imagens utilizadas para o treinamento e validação dos modelos foram obtidas em plataformas especializadas. Foram utilizadas várias técnicas para ajuste das métricas de avaliação das redes neurais. Os resultados encontrados sugerem o modelo \textit{Inception v3} como uma ferramenta valiosa para apoio ao diagnóstico de psoríase. Isso se deve ao seu desempenho satisfatório no que diz respeito à acurácia e F1-Score (97,5\% ${\pm}$ 0,2).
\end{resumo}

\section{Introdução} \label{Introduction}

Cerca de 2\% da população mundial são acometidos por lesões de pele. Uma das doenças que geram lesões na pele é a psoríase. A psoríase é uma doença autoimune que gera grande impacto negativo na qualidade de vida dos portadores. Apesar de não ser contagiosa, é crônica, inflamatória, incurável e apresenta acentuado estigma social devido à aparência das regiões afetadas \cite{oms2016relatorio}. A doença pode se manifestar nos cotovelos, joelhos e couro cabeludo até, no mais grave dos casos, o corpo inteiro \cite{sbd2012consenso}. É caracterizada pela presença de lesões róseas ou avermelhadas bem delimitadas com descamação intensa, sendo do tamanho de gotas ou moedas \cite{neto1990manual}.

O diagnóstico é feito a partir da apresentação clínica dos sinais e sintomas e do histórico do paciente. Em alguns casos é solicitada uma biópsia para fins de confirmação. O diagnóstico possui certa subjetividade, pois a análise dependerá da experiência do profissional da área que fará a avaliação \cite{vilefort2022aspectos}. O médico, porém, geralmente com base no que consegue observar em termos de sinais morfológicos, distribuição e disposição da lesão no corpo, cor, descamação, entre outros sinais, determina se a lesão é característica de psoríase ou não. A inspeção visual dos sinais de psoríase requer, por exemplo, que o médico conheça o padrão de apresentação da doença. Esse padrão de apresentação pode ser identificado através de marcadores visuais bem conhecidos. A existência de padrões de apresentação favorece o uso de técnicas de identificação automatizadas. Como consequência, o diagnóstico de psoríase pode se tornar mais objetivo e eficiente.

Em Computação não é difícil encontrar aplicações de Inteligência Artificial que oferecem suporte ao diagnóstico médico. Elas podem ser encontradas na área da Cardiologia \cite{silva2022cardiac}, Oftalmologia \cite{moura2021detecção}, Oncologia \cite{guimaraes2018using}, entre outras. Nessas áreas, Redes Neurais Convolucionais (\textit{Convolutional Neural Network} \textemdash~ CNNs) são a técnica mais utilizada para automatizar diagnósticos médicos. Por exemplo, modelos de Redes Neurais de Aprendizado Profundo permitem identificar padrões em imagens \cite{dsa_deeplearningbook}.

Nesse trabalho, são avaliados três modelos de CNNs, ResNet50 \cite{he2016deep}, \textit{Inception v3} \cite{szegedy2016rethinking} e VGG19 \cite{simonyan2014very}, para classificar imagens de lesões de peles afetadas por psoríase. Esses classificadores poderão, após validações futuras, colaborar no diagnóstico da doença.

Esse artigo está organizado da seguinte forma. A Seção \ref{RelatedWork} apresenta alguns trabalhos relacionados. A Seção \ref{Methods&Stuff} apresenta as estratégias para aquisição das imagens utilizadas no trabalho, bem como os métodos para particionamento de dados, pré-processamento das imagens, treinamento das redes neurais e as configurações definidas para garantir uma boa relação custo-benefício em relação ao desempenho dos modelos CNNs escolhidos. A Seção \ref{Results&Discussion} apresenta os resultados e discussão sobre o comportamento das redes neurais durante a realização dos experimentos. A Seção \ref{Conclusion} apresenta as conclusões e os trabalhos futuros.

\section{Trabalhos Relacionados} \label{RelatedWork}

\cite{zhao2020smart} apresenta um sistema inteligente de identificação de psoríase baseado em imagens. Esse sistema não prescinde de um dermatoscópio, instrumento muito comum em consultórios dermatológicos para avaliar lesões de pele, cabelos e unhas.

As imagens de entrada foram classificadas em nove tipos de manifestações de doenças dermatológicas: líquen plano, lúpus eritematoso, parapsoríase, carcinoma basocelular, carcinoma espinocelular, eczema, pênfigo, psoríase e ceratose seborreica. O banco de imagens utilizado foi composto por 8.021 imagens clínicas das nove classes, sendo 900 imagens de psoríase. O conjunto de dados foi dividido em 60\% para treino, 20\% para validação e 10\%  para teste. Foram utilizadas quatro arquiteturas de CNNs: DenseNet \cite{huang2017densely}, \textit{Inception v3}, InceptionResNet v2 \cite{szegedy2017inception} e Xception \cite{chollet2017xception}. O valor da área sob a curva do \textit{InceptionV3} 0,981$ {\pm}$ 0,015 foi o melhor, seguido por Xception, InceptionResNetV2 e DenseNet. Os experimentos demonstraram que o \textit{Inception v3} possuiu melhor desempenho do que as outras arquiteturas, baseado no conjunto de dados dos autores.

\cite{aggarwal2019data} apresenta uma avaliação do uso da técnica de aumento de dados na precisão do reconhecimento de imagens. As imagens de entrada se subdividem em cinco tipos de manifestações de doenças dermatológicas: acne, rosácea, psoríase, dermatite atópica e impetigo. Em geral, foram utilizadas 938 imagens, sendo 280 de psoríase. Especificamente, a psoríase antes do aumento de dados teve como resultado: 63,3\% de sensibilidade, 87,5\% de especificidade, 55,9\% de precisão e 59,4\% de \textit{F1-Score}. Depois do aumento de dados obteve-se uma melhora de 3,4\% na sensibilidade, 1,7\% na especificidade, 4,7\% na precisão e 4,1\% no \textit{F1-Score}.

\cite{pezo2020computer} apresenta um sistema para a deteção de doenças de pele a partir de imagens fotográficas. Para isso foram utilizadas redes neurais para classificar as imagens. As imagens de entrada se subdividem em dois tipos de manifestações de doenças dermatológicas: impetigo e psoríase. Em geral, 102 imagens de impetigo e 126 de psoríase, totalizando 228 imagens. Três algoritmos foram utilizados para a implementação de redes neurais pré-treinadas: \textit{Inception v3}, VGG16 \cite{simonyan2014very} e ResNet50. Foram determinadas para treinamento 1680 imagens de cada classe. Os melhores resultados foram obtidos em dez épocas. Excelentes resultados foram alcançados na deteção de doenças de pele com o algoritmo \textit{Inception v3}.

Diferentemente dos trabalhos anteriores, aqui é apresentada uma abordagem específica de classificação binária de peles com psoríase e peles saudáveis. Para determinar a arquitetura de CNN que apresenta melhor desempenho, três modelos ResNet50, Inception v3 e VGG19 são comparados. Para isso, foram utilizadas técnicas de aumento artificial de dados, reforçando seu uso somente no conjunto de treino e garantindo que a capacidade de generalização do modelo seja avaliada em dados utilizados durante os treinos e validação dos modelos, porém realistas. Além disso, propôs-se o uso da técnica de validação cruzada K-\textit{Fold} \cite{arlot2010survey} para avaliar os modelos. Isso viabiliza uma avaliação mais confiavelmente e reduz a variabilidade dos resultados \cite{bouckaert2004evaluating}. Por fim, foi utilizado um conjunto de dados públicos que permitisse a replicabilidade e o acesso menos restritivo aos dados.

\section{Materiais e Métodos} \label{Methods&Stuff}

\begin{figure} [ht]
\centering
\includegraphics[width=1\textwidth]{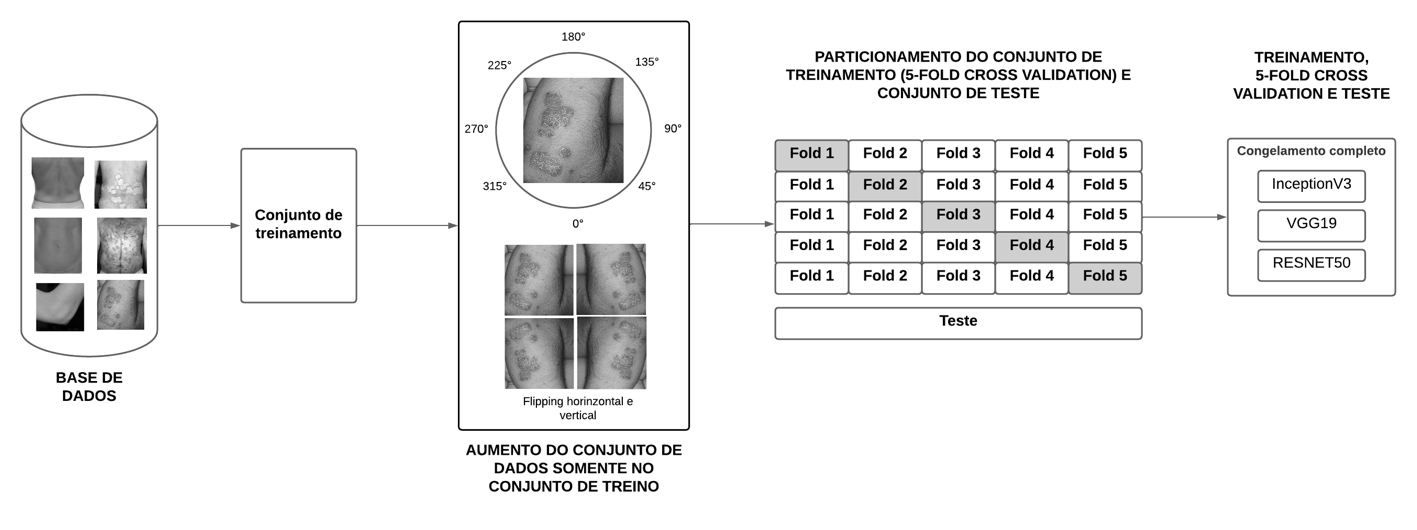}
\caption{Etapas do método proposto.}
\label{fig:metodoProposto}
\end{figure}

Escolher um classificador de imagens de psoríase mais adequado significa observar três aspectos importantes que afetam o desempenho de CNNs na classificação de imagens. Esses aspectos são a coleta e, consequentemente, a definição da base de dados, o aumento artifical de dados, o congelamento total da camada de convolução e o treinamento do classificador. Para isso, é necessário realizar um amplo estudo utilizando técnicas de validação cruzada K-\textit{Fold} e otimização dos valores dos hiper parâmetros. A Figura \ref{fig:metodoProposto} ilustra as etapas do método proposto para a escolha.

\subsection{Base de Dados}

A aquisição de imagens de pele com psoríase validadas se mostrou um grande desafio durante a fase de levantamento da base de dados de imagens. Por questões éticas, devido à grande exposição dos pacientes em imagens dermatológicas, são necessárias autorizações e consentimento expresso das pessoas fotografadas. Isso dificulta o acesso e compartilhamento dessas imagens. Foram realizados contatos com associações e especialistas da área, porém, nenhuma resposta positiva foi obtida.

Levando em conta o contexto apresentado, o conjunto de imagens da doença de pele psoríase foi coletado a partir das seguintes plataformas de dermatologia: Danderm \footnote{https://danderm-pdv.is.kkh.dk/atlas/index.html}, Atlas de Dermatologia \footnote{https://www.atlasdermatologico.com.br/}, DermatoWeb \footnote{http://dermatoweb.udl.es/}, DermNet NZ \footnote{https://dermnetnz.org/topics/psoriasis/}, DermIS \footnote{https://www.dermis.net/}, Hellenic Dermatlas \footnote{http://www.hellenicdermatlas.com/en/} e Conselho Internacional de Psoríase \footnote{https://www.psoriasiscouncil.org/imagelibrary.html}. O total de imagens obtidas foi de 1.502. Os dados desses repositórios são de domínio público e podem ser utilizados para fins educacionais. Além disso, as imagens foram validadas por dermatologistas e possuem etiquetas de diagnóstico.

As imagens com lesões localizadas em partes específicas como, por exemplo, couro cabeludo, unhas e partes íntimas foram desconsideradas, pois destoavam das demais na aparência ou eram cobertas por cabelos. Também foram desconsideradas imagens borradas e sem foco. Após esse refinamento, a quantidade de imagens com psoríase resultou em 1.130.

O conjunto de imagens de peles saudáveis, de diferentes áreas do corpo, foi obitido de dados NTU \cite{nhat2014preliminary} do Biometrics and Forensics Lab. Esse conjunto totalizou 20 imagens de regiões, como, por exemplo, costas, tórax, antebraços e pernas. Essa quantidade reduzida de imagens representa somente uma parte do conjunto de dados inteiro. Uma solicitação para a aquisição do conjunto como um todo foi enviada, porém nenhuma resposta foi obtida. Como recurso alternativo, a maior parte das imagens de pele saudável da base de dados foram coletadas do Google Images\textregistered. Esse conjunto totalizou 1.130 imagens. Por fim, o conjunto de dados levantado tem como resultado 2.260 imagens totais.

\subsection{Particionamento do Conjunto de Dados} \label{DataSetPartitioning}

O particionamento do conjunto de dados adotado nesse trabalho realizado utilizando a validação cruzada K-\textit{Fold}. Essa validação realiza múltiplas partições do conjunto total de dados em \textit{k} partições denominadas \textit{folds}. Os \textit{folds} são utilizados para treinamento e validação e alternam de forma iterativa. Dessa forma, cada parte do conjunto de dados em alguma iteração se torna o conjunto retido. Após o processamento ser concluído, para obter a estimativa de desempenho final, as \textit{k} medições são agregadas utilizando a média e o desvio padrão \cite{arlot2010survey}.

\begin{figure}[ht]
\centering
\includegraphics[width=1\textwidth]{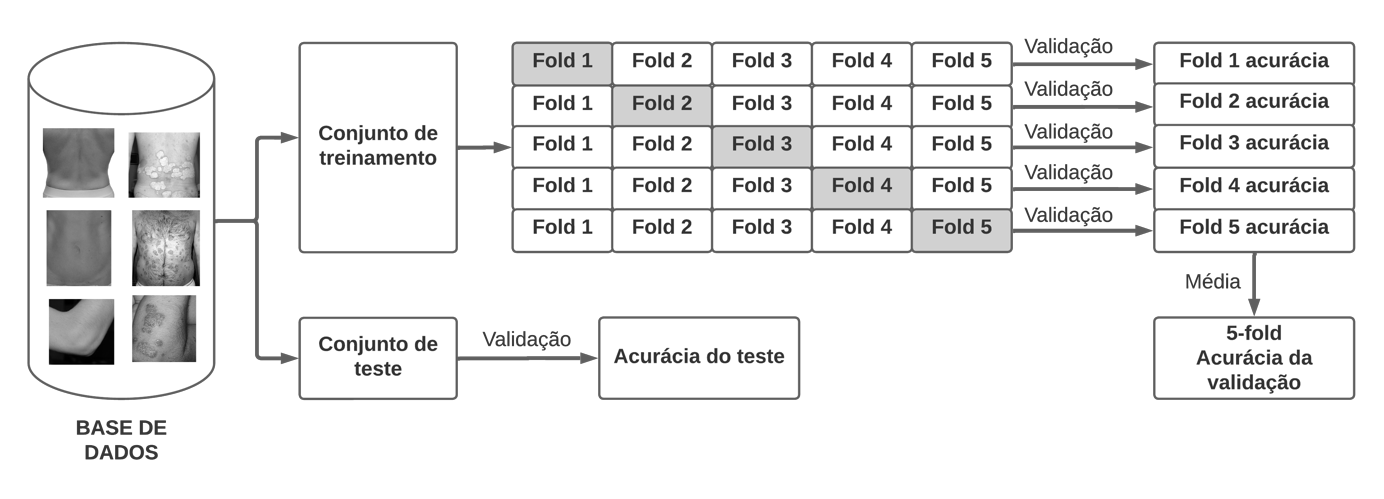}
\caption{Validação cruzada K-\textit{Fold} estratificada com \textit{k} sendo cinco.}
\label{fig:kfold}
\end{figure}

Optou-se por embaralhar e particionar dois conjuntos estratificados. Desse modo, as classes são representadas de forma semelhante em cada conjunto, de acordo com a distribuição original dos dados. Em seguida, quatro \textit{folds} foram usados para treinamento e um para teste, conforme pode ser observado na Figura \ref{fig:kfold} em que os \textit{folds} de teste estão representados de forma sombreada. A cada \textit{k} vezes em que o processo de validação é repetido, um dos \textit{folds} de treinamento é escolhido para a validação do treinamento. Assim, o modelo é avaliado em diversos conjuntos de dados, minimizando o risco de overfitting, no qual ele se ajusta demais aos dados de treinamento e falha na generalização para novos dados. Isso aumenta a confiabilidade dos resultados, tornando-os mais precisos e consistentes em um único conjunto.

\subsection{Pré-Processamento das Imagens} \label{ImagePreProcessing}

Como as amostras foram coletadas de diferentes fontes, é necessário redimensionar todas as imagens para o mesmo tamanho, de acordo com o permitido como entrada dos modelos selecionados para esse trabalho. As imagens foram redimensionadas, portanto, para 224 x 224 pixels para treinar e testar os modelos ResNet50 e VGG19 e 299 x 299 pixels para o \textit{Inception v3}. Vale ressaltar que esse processo foi realizado de forma automática com a biblioteca Keras \footnote{https://keras.io/}.

Todas as imagens foram divididas por 255 para normalizar suas intensidades de \textit{pixel} e colocá-las em uma escala entre 0 (zero) e 1 (um). Como esses valores são escalados para um intervalo menor, as imagens são melhor processadas por modelos de rede neural, tornando a computação mais eficiente. Além disso, a normalização também pode ajudar a evitar problemas de saturação em algumas funções de ativação. A divisão por 255 é utilizada especificamente porque este é o maior valor que um pixel de uma imagem em escala de cinza ou RGB pode ter.

O conjunto de dados separado para treino foi submetido ao procedimento de aumento artificial de dados. Foram realizadas rotações incrementais das imagens com valores aleatórios entre 0° (zero grau) e 20° e a realização de giro horizontal e vertical, a fim de aumentar a diversidade dos dados de treinamento e prevenir o \textit{overfitting}.

\subsection{\textit{CNNs} e Treinamento} \label{CNNsTraining}

O desafio de projetar uma CNN é um processo demorado e envolve muitas decisões de projeto. Algumas dessas decisões estão baseadas em experimentos e testes empíricos e outras na experiência e julgamento do projetista da rede, que deve contar com conhecimentos científicos e com práticas estabelecidas pela literatura. Para a realização desse trabalho optou-se por utilizar três modelos pré-treinados disponíveis na biblioteca Keras \footnote{https://keras.io/applications/}, a \textit{Inception v3}, a ResNet50 e a VGG19.

A arquitetura \textit{Inception} \cite{szegedy2016rethinking} é composta por blocos convolucionais denominados \textit{Inceptions Modules} que são empilhados de forma que suas estatísticas de correlação de saída variam e as camadas mais altas capturam as características de maior abstração. Além disso, possui filtros de resolução 1x1, 3x3 e 5x5 \cite{szegedy2016rethinking}. A VGG \cite{simonyan2014very} possui 138 milhões de parâmetros. Essa rede possui filtros de resolução 3x3. Isso demostra que é possível criar uma rede mais profunda sem aumentar tanto a quantidade de parâmetros, permitindo o aprendizado de mais não-linearidades \cite{simonyan2014very}. A ResNet \cite{he2016deep} possui 152 camadas. As camadas convolucionais utilizam a função de ativação ReLu \cite{agarap2018deep} e as camadas totalmente conectadas, a função de ativação \textit{softmax}. O grande avanço em relação à essa arquitetura deve-se não somente ao grande número de camadas, mas ao conceito de aprendizado residual. Em momentos em que a arquitetura se torna muito densa, a precisão das redes tende a diminuir. A alternativa proposta para esse problema foi a introdução de conexões entre a saída de camadas anteriores e a saída de camadas subsequentes. Ou seja, a saída de uma camada é transmitida tanto para as camadas subsequentes, quanto para as saídas das camadas posteriores \cite{he2016deep}.

A escolha das CNNs ocorreu de acordo com suas características a fim de analisar diferentes resultados em diferentes configurações, considerando também seus resultados na competição do \textit{ImageNet} \cite{russakovsky2015imagenet}. As principais características consideradas ao analisar a competição foram o tamanho da CNN, suas respectivas acurácias top-1 e top-5 no conjunto de dados de validação, a profundidade e quantidade de parâmetros de cada rede. O comum em todas é a disponibilidade em várias bibliotecas de aprendizado profundo, como o \textit{TensorFlow} \footnote{https://www.tensorflow.org/tutorials/} e o \textit{Keras}, que as torna fáceis de implementar. Outro ponto importante é a possibilidade de transferência de aprendizado que exige um conjunto de dados menor, além de reduzir o tempo e o custo de treinamento.

\subsubsection{Transferência de Aprendizado} \label{LearningTransfer}

Treinar uma rede convolucional para uma tarefa específica exige grande custo computacional e um conjunto de dados abundante. Para isso, é comum utilizar a transferência de aprendizado (\textit{Transfer Learning}). A transferência de aprendizado ocorre quando o processo de aprendizado parte de padrões aprendidos anteriormente ao resolver um problema diferente. Com isso, economiza-se tempo e recursos computacionais \cite{pan2010survey}. Há três estratégias para a utilização da transferência de aprendizado. A primeira é a utilização de modelo pré-treinado como extrator de características. A segunda é o ajuste fino do modelo pré-treinado. E a terceria é o treinamento de rede usando iniciação por transferência. No caso da primeira estratégia, o modelo é geralmente treinado em um conjunto de dados muito grande e genérico para extrair características de novas imagens. Em seguida, essas características são usadas para treinar um novo modelo com um conjunto de dados menor e mais específico. A camada final é substituída por uma camada densa que possui um número de unidades igual ao número de classes no novo conjunto de dados. Nesse trabalho, optou-se pela primeira estratégia. A opção pela primeira estratégia foi baseada nas vantagens oferecidas pela abordagem.

A utilização de um conjunto de dados pequeno para o treinamento de uma rede neural do zero pode representar um desafio, pois a falta de exemplos suficientes pode prejudicar o desempenho do modelo. Ao utilizar um modelo pré-treinado como extrator de características, é possível reutilizar o conhecimento prévio do modelo para extrair recursos das imagens e melhorar o desempenho do modelo, dessa forma, contornando a limitação do conjunto de dados. Outro ponto é a melhor generalização. Como o modelo pré-treinado em um conjunto de dados grande e diverso, a extração de características é mais robusta e abrangente. Por fim, há uma possível redução do risco de \textit{overfitting}, pois a rede neural precisa ajustar somente os pesos da camada de classificação, em vez de ajustar todos os pesos da rede, resultando em um treinamento mais fácil e eficiente.

\subsection{Configurações} \label{Setups}

Durante a fase de treinamento dos modelos ocorre o que é chamado de taxa de aprendizado (\textit{learning rate}) \cite{bottou2012stochastic}. A taxa de aprendizado é um hiperparâmetro que controla o grau de alteração do modelo em resposta ao erro estimado, cada vez que os pesos do modelo são alterados. Escolher uma taxa de aprendizado é desafiador. Se um valor muito pequeno é definido, o treinamento pode se tornar longo e, com isso, pode ocorrer travamento. Por outro lado, um valor muito grande pode resultar em um aprendizado muito rápido de um conjunto de pesos abaixo do ideal, ou ainda em um processo de treinamento instável.

Para atingir o valor mais adequado da taxa de aprendizado para os modelos escolhidos, foram utilizadas algumas técnicas, definidas na literatura, como, por exemplo, identificação do platô, avaliação de desempenho, salvamento do modelo de melhor desempenho para aplicação posterior no conjunto de dados, definição da quantidade de épocas, otimização e definição do \textit{Batch}. Essas técnicas visam acelerar a finalização do treinamento, sobretudo nos casos em que os modelos não apresentem bom desempenho.

Para identificar o platô de cada modelo utilizou-se a acurácia do conjunto de validação. O fator de redução aplicado foi de 1e-5 e a paciência (\textit{patience}) definida 5. Isso sifnifica que o modelo aguarda cinco épocas sem melhorar a métrica antes de reduzir a taxa de aprendizagem aplicando o fator. O delta mínimo foi 0,5, ou seja, a quantidade mínima de melhoria na métrica observada para que seja considerada significativa.

A avaliação do desempenho dos modelos foi feita usando a métrica da acurácia do conjunto de validação, com a paciência definida em 14 épocas. O objetivo é obter a maior acurácia possível durante o treinamento. Quando a acurácia não melhora por no máximo 14 épocas consecutivas, o trenamento é interrompido automaticamente. Isso evita que o modelo continue sendo treinado desnecessariamente. Essa técnica pode ajudar a eviter o sobreajuste (\textit{overfitting}) e melhorar a generalização do modelo.

O modelo que obteve o melhor desempenho durante o treinamento, ou seja, a maior taxa de acurácia para o conjunto de validação após todas as épocas, é salvo para que possa ser aplicado posteriormente no conjunto de testes. Ao salvar o melhor modelo, é possível garantir que o modelo mais promissor seja usado para fazer previsões em novos dados.

A quantidade de épocas (\textit{Epoch}) foi definida em no máximo 50 e para a otimização foi usado o \textit{Adam} \cite{kingma2014adam}, que possui uma otimização computacionalmente eficiente. Além disso, requer pouca memória e pouco ajuste de hiperparâmetros. Essa escolha é muito popular em problemas de \textit{Machine Learning} e \textit{Deep Learning} \cite{kingma2014adam}.

O \textit{Batch} foi definido em 32, ou seja, o algoritmo treina os modelos por partes, em blocos de 32 amostras, até que todo o conjunto de treinamento seja utilizado.

Devido à natureza do problema proposto ser de classificação e envolver uma solução que tem como propósito ser uma ferramenta de apoio aos profissionais de saúde na confirmação de hipóteses diagnósticas, identificou-se a relevância da métrica de acurácia sobre a perda. A acurácia é particularmente importante nesse cenário em que se deseja minimizar o número de erros de classificação, pois a classificação incorreta pode ter consequências graves. Um alto valor de acurácia pode ser um indicador importante de que o modelo está realizando corretamente a tarefa para a qual foi projetado.

Por fim, a configuração das redes pré-treinadas é de extrema importância, pois elas foram implementadas para solucionar o problema do \textit{ImageNet}, ou seja, todas possuem como saída 1.000 classes. Torna-se necessário, portanto, adaptá-las para o cenário desse trabalho, que possui um conjunto de dados com apenas duas classes. Para isso, as redes são baixadas com ausência do último bloco de camadas, pois é nele que ocorre o processo de classificação. Em seguida, é adicionada uma camada \textit{GlobalAveragePooling2D} \cite{keras-globalaveragepooling2d} para reduzir a dimensionalidade dos dados para uma dimensão e, por se tratar de uma classificação binária, é necessário reduzir para uma camada densa com ativação sigmoide com uma unidade de saída. Por último, a rede é compilada com a nova configuração.

\subsection{Métricas de Avaliação} \label{Evaluation}

As métricas utilizadas para avaliar as redes definidas foram: acurácia, precisão, revocação, especificidade, \textit{F1-Score} e curva de ROC. Essas métricas fazem uso de uma matriz de confusão, que é uma matriz quadrada e visa comparar os valores previstos pelo modelo com os valores esperados (verdadeiros). A diagonal principal contém os acertos do modelo e as posições restantes denotam erros. Há quatro tipos de rótulos em uma matriz de confusão. Verdadeiro negativo (TN) representa o número de amostras rotuladas corretamente para a classe negativa. Verdadeiro positivo (TP) indica o número de amostras rotuladas corretamente para a classe positiva. Falso negativo (FN) é número de amostras rotuladas incorretamente para a classe positiva. E Falso positivo (FP) é o número de amostras rotuladas incorretamente para a classe negativa.

A acurácia, ou seja, a representação das previsões corretas de um modelo, é calculada conforme a Equação \ref{eq:acuracia}. A precisão foi calculada conforme a Equação \ref{eq:precisao}. A sensibilidade, ou revocação (\textit{recall}), que avalia todas as amostras positivas rotuladas corretamente, foi calculada conforme a Equação \ref{eq:recall}. A especificidade, utilizada para avaliar todas as amostras negativas rotuladas  corretamente, é descrita conforme a Equação \ref{eq:especificidade}. O \textit{F1-Score} é dado pela média harmônica da precisão e a sensibilidade, descrita conforme a Equação \ref{eq:f1score}.

\begin{equation}
 Acuracia = \frac{TP + TN}{TP + TN + FP + FN}\label{eq:acuracia}
\end{equation}

\begin{equation}
 Precisao = \frac{TP}{TP + FP}\label{eq:precisao}
\end{equation}

\begin{equation}
 Sensibilidade = \frac{TP}{TP + FN}\label{eq:recall}
\end{equation}

\begin{equation}
 Especificidade = \frac{TN}{TN + FP}\label{eq:especificidade}
\end{equation}

\begin{equation}
 F1-Score = \frac{2 * Precisao * Sensibilidade}{Precisao + Sensibilidade}\label{eq:f1score}
\end{equation}

Mesmo com essas métricas, é possível que ocorra alguma dificuldade de avaliação ao comparar os modelos. Essa dificuldade ocorre quando, por exemplo, se trata de classes distribuídas de forma assimétrica e quando há erros de classificação com custos diferentes \cite{fawcett2006introduction}. Para isso, costumasse utilizar a curva Característica de Operação do Receptor (\textit{Receiver Operating Curve} \textemdash~ ROC). A curva ROC é um gráfico que relaciona a sensibilidade com a especificidade no experimento, muito utilizada com classificadores binários.

Existem métricas que simplificam a visualização dos desempenhos dos modelos a partir da curva ROC. A mais utilizada é a área sob a curva (\textit{Area Under the Curve} \textemdash~ AUC). Um desempenho ótimo é caracterizado pela AUC se aproximando de 1, enquanto o desempenho 0.5 significa que o modelo não está distinguindo as classes corretamente. Por ser bastante utilizada em classificadores de patologias na área médica \cite{swets1988measuring}, essa métrica é utilizada nesse trabalho.

\section{Resultados e Discussão} \label{Results&Discussion}

Os resultados obtidos nesse trabalho destacam quanto o aumento artificial de dados, o congelamento total da camada de convolução e o treinamento impactam no classificador. Nessa seção os resultados são apresentados e é discutido uso das estratégias de treinamento utilizadas e avaliado o desempenho da classificação de cada CNN, segundo as métricas descritas nas Seções \ref{Setups} e \ref{Evaluation}.

A Tabela \ref{fig:afericaoMetricasComSemAumento} apresenta as medidas de desempenho dos três modelos (ResNet50, VGG19 e \textit{Inception v3}) com e sem a aplicação da técnica de aumento de dados, em termos de média de acurácia, precisão, \textit{recall}, especificidade e \textit{F1-score} de validação cruzada com 5-\textit{Fold}.

\begin{table}[ht]
\centering
\caption{Métricas gerais em termos de média e desvio padrão.}
\label{fig:afericaoMetricasComSemAumento}
\includegraphics[width=1\textwidth]{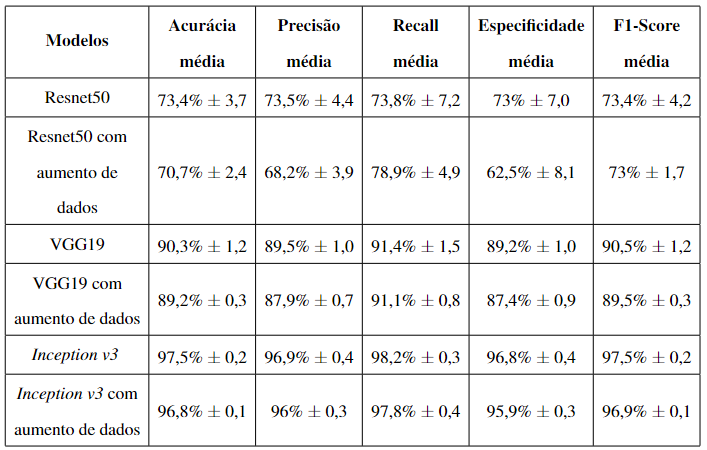}
\end{table}

Comparando os modelos sem aumento de dados, o ResNet50 apresenta a menor acurácia média (73,4\%) e as menores métricas de precisão (73,5\%) e \textit{recall} (73,8\%). O \textit{Inception v3}, por outro lado, apresenta a maior acurácia média (97,5\%) e as maiores métricas de precisão (96,9\%) e \textit{recall} (98,2\%). O \textit{Inception v3} apresenta a maior especificidade média (96,8\%), enquanto o ResNet50 apresenta a menor especificidade média (73\%).

Ao adicionar aumento de dados, observa-se uma queda na acurácia média do ResNet50 (70,7\%), com métricas de precisão (68,2\%) e especificidade (62,5\%) ainda menores. O VGG19 com aumento de dados, por sua vez, apresenta uma acurácia média ligeiramente menor (89,2\%) em comparação com o modelo sem aumento de dados, apresentando essa ligeira queda em todas as outras métricas. O \textit{Inception v3} com aumento de dados, porém, apresenta uma acurácia média semelhante (96,8\%) ao modelo sem aumento de dados, mas com desempenho ligeiramente menor em precisão (96\%) e especificidade (95,9\%).

É importante observar o desvio padrão (DP) das métricas apresentadas. O ResNet50, que apresentou os maiores valores de DP sem o aumento de dados, mostrou uma queda significativa após o aumento de dados, no que diz respeiro à métrica de \textit{recall}, com uma diferença de 2,3 pontos. Essa diminuição do DP ocorreu em todas as métricas de todos os modelos, mostrando o impacto da regularização do conjunto de treino na estabilidade dos modelos.

Em geral, o modelo \textit{Inception v3} apresenta o melhor desempenho em todas as métricas, com e sem aumento de dados, seguido pelo VGG19. O ResNet50 apresenta o desempenho mais baixo, especialmente com aumento de dados, com desempenho significativamente pior em comparação com os outros modelos.

Vale ressaltar que o aumento de dados é uma técnica de regularização da fase de treinamento de um modelo. Ele consiste em aumentar a quantidade de dados disponíveis para treino, gerando novas imagens a partir das imagens existentes por meio de transformações, como, por exemplo, rotação, ampliação e corte. Como é técnica muito útil quando há poucos dados disponíveis, permite que o modelo processe mais exemplos de imagens e, consequentemente, aprenda melhor os padrões presentes nos dados, generalizando e evitando o \textit{overfitting}. No entanto, o aumento de dados nem sempre apresenta impacto positivo significativo nas métricas de desempenho do modelo, como foi possível observar nos resultados dessa pesquisa. A diminuição dos valores de desvio padrão comprovou a robustez dos modelos treinados, o que permitiu conseguir a regularização, que é objetivo da técnica de aumento de dados.

\subsection{Discussão} \label{Discussion}

A principal característica observada durante os experimentos é o desempenho superior do modelo \textit{Inception v3} em comparação aos outros dois modelos no contexto desse trabalho. Isso foi evidenciado em todas as métricas de desempenho e melhor visualizado na curva ROC e na matriz de confusão.

Existem alguns fatores que podem ter contribuído para a \textit{Inception v3} apresentar um desempenho melhor em relação aos outros modelos avaliados. O principal deles é a relação entre a arquitetura do modelo e a complexidade do conjunto de dados. Nesse trabalho, as imagens dermatológicas requerem uma análise detalhada da textura e da cor da pele, bem como a identificação de bordas e fronteiras das lesões. Isso pode ser particularmente difícil em lesões que são sutis ou que ocorrem em áreas de pele com dobras ou textura irregular, como é o caso da psoríase. Além disso, há também a variabilidade na aparência das lesões entre indivíduos, a presença de vários tipos de lesões em uma única imagem, a influência de condições variáveis de iluminação e a presença de artefatos e ruídos. A \textit{Inception v3} é uma CNN com uma arquitetura mais complexa e profunda do que a ResNet50 e a VGG19. Enquanto a VGG19 possui 19 camadas convolucionais e a ResNet50 possui 50 camadas, a \textit{Inception v3} tem 159. Ela foi projetada para lidar com imagens complexas e possui camadas de convolução profundas que podem extrair recursos mais precisos das imagens.

Um segundo ponto observado foi a eficácia da técnica de aumento de dados nos modelos estudados. Apesar de não ter apresentado um impacto positivo significativo nas métricas de desempenho, houve uma diminuição do desvio padrão em todos os modelos. Um exemplo é o desvio padrão da métrica de \textit{recall}, que apresentou uma queda de 2,3 pontos após a aplicação do aumento de dados. Isso comprovou a robustez dos modelos treinados.

\begin{figure}[ht]
\centering
\begin{minipage}{.5\textwidth}
  \centering
  \includegraphics[width=\linewidth]{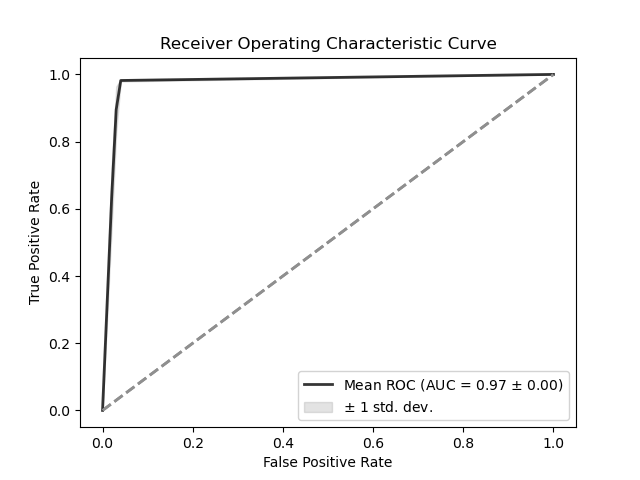}
\end{minipage}\hfill
\begin{minipage}{.5\textwidth}
  \centering
  \includegraphics[width=\linewidth]{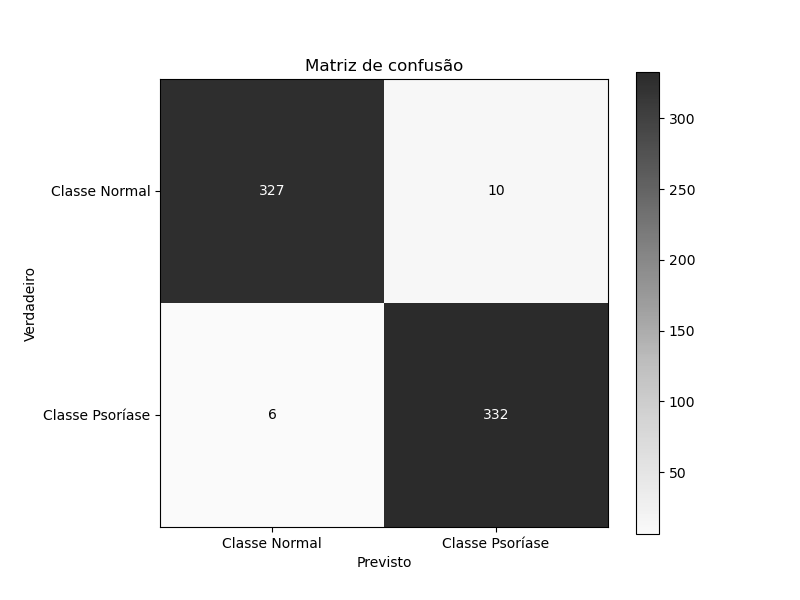}
\end{minipage}
\caption{Curva ROC e Matriz de Confusão do modelo \textit{Inception v3}.}
\label{fig:rocInception}
\end{figure}

As métricas foram avaliadas de forma geral. Entretanto, uma vez que o experimento com o modelo \textit{Inception v3} sem aumento de dados obteve resultados superiores ao experimento com aumento de dados, optou-se por analisar seus resultados com destaque. Ao observar a matriz de confusão da Figura \ref{fig:rocInception}, observou-se, por exemplo, que a média do número de casos corretamente classificados como psoríase foi 332. A média do número de casos corretamente classificados como saudável foi 327 e a média do número de casos incorretamente classificados como psoríase foi 10. Nesse caso, os erros configuram-se do tipo 1. Por fim, a média do número de casos incorretamente classificados como saudável foi 6. Nesse caso, os erros configuram-se do tipo 2.

Ao analisar modelos para auxílio à triagem médica, os erros podem ter diferentes pesos. No caso desse trabalho, os erros do tipo 2 são mais graves, pois significam que o paciente pode ser diagnosticado erroneamente como não tendo psoríase, quando na verdade ele tem. Isso pode levar a atrasos no tratamento e agravar potencialmente a condição médica do paciente. Levando em consideração que o modelo escolhido possui seis ocorrências de erro do tipo 2 e um \textit{F1-Score} de 97,5\%, o que significa que o modelo é capaz de identificar corretamente a maioria das instâncias positivas (\textit{recall} alto), ao mesmo tempo que minimiza o número de falsos positivos (precisão alta), pode-se afirmar que o modelo possui resultados satisfatórios. Isso evidencia a capacidade do modelo em lidar com a classificação correta das imagens.

Tendo em vista que a deteção precoce da psoríase tem um impacto positivo significativo na vida dos pacientes e no sistema de saúde em geral, um modelo promissor de deteção da doença contribui para o avanço da tecnologia nessa área de pesquisa, além de trazer diversos benefícios, como, por exemplo, apoio ao diagnóstico rápido e preciso, melhoria na qualidade de vida dos pacientes, redução de custos e maior conscientização sobre a doença.

\section{Conclusões} \label{Conclusion}

Esse trabalho teve como objetivo realizar um estudo comparativo do desempenho de diferentes arquiteturas de Redes Neurais Convolucionais para classificação de imagens para a deteção de lesões afetadas pela doença de pele psoríase. Foram utilizadas três redes neurais convolucionais diferentes: ResNet50, \textit{Inception v3} e VGG19. Cada modelo foi treinado e validado utilizando a técnica de validação cruzada 5-\textit{Fold}. Os modelos foram avaliados em diferentes conjuntos de dados, reduzindo a chance de \textit{overfitting} em um único conjunto e aumentando a confiabilidade dos resultados, tornando-os mais precisos e consistentes.

No decorrer do trabalho, observou-se quanto as CNNs podem ser eficazes para reconhecer padrões valendo-se da aplicação de filtros e métodos previamente aprendidos. Utilizando os modelos pré-treinados como extratores de características, mesmo com um conjunto de imagens limitado, foi possível atingir resultados satisfatórios. Foi também possível constatar a notoriedade do uso da arquitetura \textit{Inception v3} na tarefa de classificação de imagens. O modelo atingiu as métricas de \textit{F1-Score} e acurácia em 97,5\%, evidenciando que é capaz de identificar corretamente a maioria das instâncias positivas, ao mesmo tempo que minimiza o número de falsos positivos, resultados considerados promissores para a pesquisa.

Um ponto positivo do uso da arquitetura \textit{Inception v3} para classificação de imagens da doença da pele psoríase é que forneceu resultados úteis para futuros projetos e pesquisas na área de apoio ao diagnóstico médico. Especificamente, a \textit{Inception v3} é uma escolha bem-sucedida para essa tarefa e futuras pesquisas podem usá-la como referência para comparar o desempenho de outros modelos. Com isso, tempo e recursos podem ser economizados, além de permitir uma comparação mais precisa entre diferentes abordagens. Além disso, um modelo com resultados promissores pode trazer alguns benefícios para a sociedade, como, por exemplo, a identificação precoce da doença, pois modelo ajuda a identificar a doença em estágios iniciais, o que pode melhorar as chances de recuperação e tratamento bem-sucedido; redução do tempo de diagnóstico, permitindo mais eficiência, eficácia e efetividade do tratamento; redução de custos no tratamento de doenças, pois quanto mais cedo a doença é identificada, menos recursos são necessários para tratá-la; e um melhor prognóstico, o que permite ao paciente planejar suas vidas para conseguir uma melhor qualidade de vida.

Esse trabalho não se propõe a automatizar totalmente o processo de avaliação e diagnóstico, mas sim oferecer aos profissionais de saúde uma ferramenta de apoio para a investigação de hipóteses diagnósticas. Com isso, deseja-se colaborar com o curso de tratamento e tornar as avaliações reproduzíveis e assertivas. Entretanto, é importante enfatizar que são necessários mais testes e validações para que o modelo seja utilizado em contextos reais.

Como trabalhos futuros pretende-se explorar o uso de outros hiperparâmetros, como, por exemplo, o otimizador RMSProp \cite{ruder2016overview}, e de técnicas de otimização de de hiperparâmetros, como, por exemplo, a busca em grade ou Busca Bayesiana \cite{snoek2012practical}. Além disso, planeja-se investigar o uso de outras estratégias de aumento de dados, como, por exemplo, a aplicação de filtros de brilho e contraste com objetivo de melhorar o desempenho dos modelos. Pretende-se também utilizar outras estratégias de transferência de aprendizado para fazer o ajuste fino do modelo pré-treinado. Por fim, aumentar o raio de abrangência do modelo, contemplando mais doenças de pele.

\section*{Agradecimentos}

Os autores agradecem o apoio material provido pelo Laboratório de Sistemas Inteligentes (LSI) da Universidade do Estado do Amazonas (UEA).

\bibliographystyle{sbc}
\bibliography{eniac2023}

\begin{thebibliography}{}

\bibitem[Academy 2022]{dsa_deeplearningbook}
Academy, D.~S. (2022).
\newblock Deep learning book.
\newblock \url{https://www.deeplearningbook.com.br/}.
\newblock Acessado em: 01 de outubro de 2022.

\bibitem[Agarap 2018]{agarap2018deep}
Agarap, A.~F. (2018).
\newblock Deep learning using rectified linear units (relu).
\newblock {\em arXiv preprint arXiv:1803.08375}.
\newblock DOI: 10.48550/arXiv.1803.08375.

\bibitem[Aggarwal 2019]{aggarwal2019data}
Aggarwal, S. L.~P. (2019).
\newblock Data augmentation in dermatology image recognition using machine
  learning.
\newblock {\em Skin Research and Technology}, 25(6):815--820.

\bibitem[Arlot and Celisse 2010]{arlot2010survey}
Arlot, S. and Celisse, A. (2010).
\newblock A survey of cross-validation procedures for model selection.
\newblock {\em Statist. Surv.}, 4:40--79.
\newblock First available in Project Euclid: 9 March 2010. DOI:
  10.1214/09-SS054.

\bibitem[Bottou 2012]{bottou2012stochastic}
Bottou, L. (2012).
\newblock Stochastic gradient descent tricks.
\newblock In {\em Neural Networks: Tricks of the Trade: Second Edition}, pages
  421--436. Springer.
\newblock DOI: 10.1007/978-3-642-35289-8\_25.

\bibitem[Bouckaert and Frank 2004]{bouckaert2004evaluating}
Bouckaert, R.~R. and Frank, E. (2004).
\newblock Evaluating the replicability of significance tests for comparing
  learning algorithms.
\newblock In {\em Pacific-Asia conference on knowledge discovery and data
  mining}, pages 3--12. Springer.
\newblock DOI: 10.1007/978-3-540-24775-3\_3.

\bibitem[Chollet 2017]{chollet2017xception}
Chollet, F. (2017).
\newblock Xception: Deep learning with depthwise separable convolutions.
\newblock In {\em Proceedings of the IEEE conference on computer vision and
  pattern recognition}, pages 1251--1258.
\newblock DOI: 10.48550/arXiv.1610.02357.

\bibitem[de~Dermatologia 2012]{sbd2012consenso}
de~Dermatologia, S.~B. (2012).
\newblock {\em Consenso Brasileiro de Psor{\'i}ase 2012. Guias de
  avalia{\c{c}}{~a}o e tratamento Sociedade Brasileira de Dermatologia}.
\newblock Sociedade Brasileira de Dermatologia, 2nd edition.
\newblock Acessado em: 17 de agosto de 2022.

\bibitem[Fawcett 2006]{fawcett2006introduction}
Fawcett, T. (2006).
\newblock An introduction to roc analysis.
\newblock {\em Pattern Recognition Letters}, 27(8):861--874.
\newblock DOI: 10.1016/j.patrec.2005.10.010.

\bibitem[Guimaraes et~al. 2018]{guimaraes2018using}
Guimaraes, A.~J., Ara{\'u}jo, V.~J., de~Oliveira~Batista, L., Souza, P. V.~C.,
  Ara{\'u}jo, V., and Rezende, T.~S. (2018).
\newblock Using fuzzy neural networks to improve prediction of expert systems
  for detection of breast cancer.
\newblock In {\em Anais do XV Encontro Nacional de Intelig{\^e}ncia Artificial
  e Computacional}, pages 799--810. SBC.
\newblock DOI: 10.5753/eniac.2018.4468.

\bibitem[He et~al. 2016]{he2016deep}
He, K., Zhang, X., Ren, S., and Sun, J. (2016).
\newblock Deep residual learning for image recognition.
\newblock In {\em Proceedings of the IEEE conference on computer vision and
  pattern recognition}, pages 770--778.
\newblock DOI: 10.1109/CVPR.2016.90.

\bibitem[Huang et~al. 2017]{huang2017densely}
Huang, G., Liu, Z., Van Der~Maaten, L., and Weinberger, K.~Q. (2017).
\newblock Densely connected convolutional networks.
\newblock In {\em Proceedings of the IEEE conference on computer vision and
  pattern recognition}, pages 4700--4708.
\newblock DOI: 10.48550/arXiv.1608.06993.

\bibitem[Keras 2023]{keras-globalaveragepooling2d}
Keras (2023).
\newblock Globalaveragepooling2d layer.
\newblock
  \url{https://keras.io/api/layers/pooling_layers/global_average_pooling2d/}.
\newblock Acessado em: 01 de outubro de 2022.

\bibitem[Kingma and Ba 2014]{kingma2014adam}
Kingma, D.~P. and Ba, J. (2014).
\newblock Adam: A method for stochastic optimization.
\newblock {\em arXiv preprint arXiv:1412.6980}.
\newblock DOI: 10.48550/arXiv.1412.6980.

\bibitem[Moura et~al. 2021]{moura2021detecção}
Moura, L. R.~d., Coelho, A.~M., and Baffa, M. d. F.~O. (2021).
\newblock Detecção automática de anomalias oculares utilizando redes neurais
  convolucionais.
\newblock In {\em Anais do VIII Encontro Nacional de Computação dos
  Institutos Federais}, pages 73--79, [S.l.]. SBC.
\newblock DOI: 10.5753/encompif.2021.15953.

\bibitem[Neto et~al. 1990]{neto1990manual}
Neto, C.~F., Cuc{\'e}, L.~C., and dos Reis, V. M.~S. (1990).
\newblock {\em Manual de dermatologia}.
\newblock Editora Manole.

\bibitem[Nhat et~al. 2014]{nhat2014preliminary}
Nhat, H. et~al. (2014).
\newblock A preliminary report on a full-body imaging system for effectively
  collecting and processing biometric traits of prisoners.
\newblock In {\em IEEE Symposium Series on Computational Intelligence}.
\newblock DOI: 10.1109/CIBIM.2014.7015459.

\bibitem[OMS 2016]{oms2016relatorio}
OMS (2016).
\newblock Relatório global sobre a psoríase.
\newblock
  \url{https://apps.who.int/iris/bitstream/handle/10665/204417/9789241565189-por.pdf}.
\newblock Acessado em: 20 de agosto de 2022.

\bibitem[Pan and Yang 2010]{pan2010survey}
Pan, S.~J. and Yang, Q. (2010).
\newblock A survey on transfer learning.
\newblock {\em IEEE Transactions on knowledge and data engineering},
  22(10):1345--1359.
\newblock DOI: 10.1109/TKDE.2009.191.

\bibitem[Pezo et~al. 2020]{pezo2020computer}
Pezo, A.~R., Nogueira, K., Bologna, R., and Montoya, G. (2020).
\newblock Computer application for the detection of skin diseases in
  photographic images using convolutional neural networks.
\newblock In {\em Latin American High Performance Computing Conference}, pages
  193--204. Springer.
\newblock DOI: 10.1007/978-3-030-68035-0\_14.

\bibitem[Ruder 2016]{ruder2016overview}
Ruder, S. (2016).
\newblock An overview of gradient descent optimization algorithms.
\newblock {\em arXiv preprint arXiv:1609.04747}.
\newblock DOI: 10.48550/arXiv.1609.04747.

\bibitem[Russakovsky et~al. 2015]{russakovsky2015imagenet}
Russakovsky, O., Deng, J., Su, H., Krause, J., Satheesh, S., Ma, S., Huang, Z.,
  Karpathy, A., Khosla, A., Bernstein, M., et~al. (2015).
\newblock Imagenet large scale visual recognition challenge.
\newblock {\em International journal of computer vision}, 115:211--252.
\newblock DOI: 10.48550/arXiv.1409.0575.

\bibitem[Silva et~al. 2022]{silva2022cardiac}
Silva, G. et~al. (2022).
\newblock Cardiac arrhythmia detection in ecg signals using graph convolutional
  network.
\newblock In {\em Anais do XXII Simpósio Brasileiro de Computação Aplicada
  à Saúde}, pages 25--35.
\newblock DOI: 10.5753/sbcas.2022.222434.

\bibitem[Simonyan and Zisserman 2014]{simonyan2014very}
Simonyan, K. and Zisserman, A. (2014).
\newblock Very deep convolutional networks for large-scale image recognition.
\newblock {\em arXiv preprint arXiv:1409.1556}.
\newblock DOI: 10.48550/arXiv.1409.1556.

\bibitem[Snoek et~al. 2012]{snoek2012practical}
Snoek, J., Larochelle, H., and Adams, R.~P. (2012).
\newblock Practical bayesian optimization of machine learning algorithms.
\newblock In {\em Advances in neural information processing systems},
  volume~25.
\newblock DOI: 10.48550/arXiv.1206.2944.

\bibitem[Swets 1988]{swets1988measuring}
Swets, J.~A. (1988).
\newblock Measuring the accuracy of diagnostic systems.
\newblock {\em Science}, 240(4857):1285--1293.
\newblock DOI: 10.1126/science.3287615.

\bibitem[Szegedy et~al. 2017]{szegedy2017inception}
Szegedy, C., Ioffe, S., Vanhoucke, V., and Alemi, A. (2017).
\newblock Inception-v4, inception-resnet and the impact of residual connections
  on learning.
\newblock In {\em Proceedings of the AAAI conference on artificial
  intelligence}, volume~31.
\newblock DOI: 10.48550/arXiv.1602.07261.

\bibitem[Szegedy et~al. 2016]{szegedy2016rethinking}
Szegedy, C., Vanhoucke, V., Ioffe, S., Shlens, J., and Wojna, Z. (2016).
\newblock Rethinking the inception architecture for computer vision.
\newblock In {\em Proceedings of the IEEE conference on computer vision and
  pattern recognition}, pages 2818--2826. IEEE.
\newblock DOI: 10.48550/arXiv.1512.00567.

\bibitem[Vilefort et~al. 2022]{vilefort2022aspectos}
Vilefort, L.~A., Souza, H., Vilela, L.~C., Tanaka, V. Y.~T., Vianna, R.~M.,
  de~S{\'a}, Y.~A., Lisb{\^o}a, A. C.~C., Oliveira, M.~S., Duarte, A. C.~S.,
  Rezende, O. G.~M., et~al. (2022).
\newblock Aspectos gerais da psor{\'\i}ase: revis{\~a}o narrativa.
\newblock {\em Revista Eletr{\^o}nica Acervo Cient{\'\i}fico},
  42:e10310--e10310.
\newblock DOI: 10.25248/reac.e10310.2022.

\bibitem[Zhao et~al. 2020]{zhao2020smart}
Zhao, S., Yang, X., Xu, W., Zhang, Y., Huang, L., Chen, W., Wei, H., and Jiang,
  B. (2020).
\newblock Smart identification of psoriasis by images using convolutional
  neural networks: a case study in china.
\newblock {\em Journal of the European Academy of Dermatology and Venereology},
  34(3):518--524.
\newblock DOI: 10.1111/jdv.15965.

\end{thebibliography}

\end{document}